\documentclass[english,aps,floatfix,12pt]{revtex4}
\usepackage{epsfig,amssymb,amsfonts,bm,psfrag}

\usepackage[active]{srcltx}

\makeatletter

\usepackage{babel}
\makeatother
\begin{document}

\def\be{\begin{equation}}
\def\ee{\end{equation}}

\title{Exploring correlations in the stochastic Yang--Mills vacuum}

\author{Dmitri Antonov}
\affiliation{Departamento de F\'isica and Centro de F\'isica das Interac\c{c}\~oes Fundamentais,
Instituto Superior T\'ecnico, UT Lisboa, Av. Rovisco Pais, 1049-001 Lisboa, Portugal}

\begin{abstract}
The correlation lengths of nonperturbative-nonconfining
and confining stochastic background Yang--Mills fields are obtained by means of a direct analytic path-integral evaluation of the Green functions of the so-called one- and two-gluon gluelumps. Numerically, these lengths turn out to be in a good agreement with those known from the earlier, Hamiltonian, treatment of such Green functions.
It is also demonstrated that the correlation function of nonperturbative-nonconfining fields decreases with the 
deviation of the path in this correlation function from the straight-line one.
\end{abstract}

\maketitle

\section{Introduction}
Stochastic vacuum model~\cite{svm} (for reviews see~\cite{rev1, rev2}) 
puts in practice the idea that it is the stochasticity  
of nonperturbative background Yang--Mills fields that provides confinement. This idea is implemented
by assuming the ensemble of fields to be predominantly Gaussian, with the two-point 
correlation function containing the part essential for confinement. Despite the numerical support by 
various lattice simulations~\cite{rev2, lat1, dm1, lat2}, this model, appealing by its simplicity, requires also 
theoretical support in the form of analytic studies of the said two-point function. Such calculations have been performed
in various Abelian models, where confinement is provided by the monopole condensation~\cite{ab}, in the instanton-vacuum model~\cite{inst},
and recently also in AdS/QCD~\cite{and}. 
In QCD itself, the first progress has been reached in Ref.~\cite{fi},
where the correlation length of the two-point 
function has been related to the gluon condensate. Furthermore, Ref.~\cite{gl} explored the possibility for 
nonperturbative-nonconfining and confining background fields 
to have different correlation lengths. Such a possibility is specific for QCD, and unlikely to exist 
in Abelian models with confinement.
This phenomenon has first been noticed in the lattice QCD simulations~\cite{lat2}, gaining  
its further analytic 
support within the theory of so-called gluelumps~\cite{gl, 3, 33}.

Gluelumps can be thought as bound states of gluons in the field of a hypothetical infinitely heavy adjoint source~\cite{cm}. In Yang--Mills theory, they define the correlation lengths of the field-strengths'
two-point function in the same way as, in full QCD, physically existing
heavy-light mesons define the correlation length of a 
nonlocal quark condensate $\left<\bar\psi(x)\Phi_{xx'}\psi(x')\right>$ \cite{gl, dejm} (here 
$\Phi_{xx'}$ is a phase factor along the straight-line path, which is provided by the heavy-quark propagator). Unlike the fundamental case, 
the adjoint case allows for two different types of heavy-light objects --- 
those with a single gluon, called one-gluon gluelumps, and those with two gluons, called two-gluon gluelumps
(cf. Fig.~\ref{Fig12} below). The first case 
is similar to the above-mentioned nonlocal quark condensate, whereas
the second case is conceptually 
different, as it corresponds to two gluons connected together with the heavy source by three fundamental strings. Specifically, it has been shown 
in Ref.~\cite{3} that the Green functions of one- and two-gluon gluelumps define respectively the 
correlation lengths of nonperturbative-nonconfining and confining stochastic Yang--Mills fields. 
Furthermore, these Green functions have been 
explored in Ref.~\cite{3}
by using respectively one- and two-body relativistic Hamiltonians with linear potentials.

In the next Section, we accomplish the full calculation of 
quantum-mechanical path integrals representing those Green functions, which has been started in Ref.~\cite{33}. 
This analysis will in particular allow us to
find the above-mentioned two vacuum correlation lengths, and compare them with those 
obtained in Ref.~\cite{3} within the Hamiltonian approach. Such a calculation turns out to be possible 
by virtue of an effective parametrization of minimal areas swept out by the strings in the gluelumps.
This parametrization, suggested in Ref.~\cite{4}, has been successfully used there to account 
for confinement of virtual gluons in the polarization operator.
In Section~III, we summarize the main results of the paper. 
In Appendix~A, we calculate the Green function of a one-gluon gluelump for paths deviating from the 
straight-line one, and show that the corresponding correlation function of nonperturbative-nonconfining background fields decreases with the deformation of the path.

\section{Analytic calculation of the vacuum correlation lengths}

Stochastic vacuum model suggests the following parametrization for the two-point correlation function of gluonic field strengths~\cite{svm}:
$$\frac{g^2}{N_c}{\,}{\rm tr}{\,}\left<F_{\mu\nu}^a(x)T^a\Phi_{xx'}F_{\lambda\rho}^b(x')T^b\Phi_{x'x}\right>=
(\delta_{\mu\lambda}\delta_{\nu\rho}-\delta_{\mu\rho}\delta_{\nu\lambda})D(x-x')+$$
\be
\label{09}
+\frac12\left[\partial_\mu^x\left((x-x')_\lambda\delta_{\nu\rho}-(x-x')_\rho\delta_{\nu\lambda}\right)+
\partial_\nu^x\left((x-x')_\rho\delta_{\mu\lambda}-(x-x')_\lambda\delta_{\mu\rho}\right)\right]D_1(x-x').
\ee
Here, $\Phi_{xx'}\equiv {\cal P}{\,}\exp\left[ig\int_{x'}^{x}dz_\mu A_\mu^a(z)T^a\right]$ is a phase 
factor along some path interconnecting the points $x'$ and $x$.
For the rest of the present Section, this path is chosen along the straight line. Furthermore, $T^a$'s
are the SU($N_c$) generators in the fundamental representation, the average $\left<\ldots\right>$
is taken with respect to the Yang--Mills action $\frac14\int d^4x(F_{\mu\nu}^a)^2$, 
$F_{\mu\nu}^a=\partial_\mu A_\nu^a-\partial_\nu A_\mu^a+gf^{abc}A_\mu^b A_\nu^c$ is the Yang--Mills field-strength tensor, $a=1, \ldots, N_c^2-1$. The functions $D_1(x)$ and $D(x)$ parametrize respectively
the nonperturbative-nonconfining and confining self-interactions of stochastic background fields. 
In this Section, we
calculate $D_1(x)$ and $D(x)$ analytically, using their relation to the infra-red Green functions of 
one- and two-gluon gluelumps~\cite{3}: 
\begin{equation}
\label{D}
D_1(x)=-4g^2C_{\rm f}\frac{dG(x)}{dx^2},~~~~
D(x)=\frac{g^4(N_c^2-1)}{2}S(x),
\end{equation}
where $C_{\rm f}=(N_c^2-1)/(2N_c)$ is the quadratic Casimir operator of the fundamental representation. 
Equations~(\ref{D}) hold at large distances, $|x|\ge{\cal O}(\sigma_{\rm f}^{-1/2})$, where $\sigma_{\rm f}$ is the string tension in the fundamental representation~\cite{3}. In this infra-red regime, perturbative contributions are 
negligible, and the Wilson-loop averages appearing in the Green functions $G(x)$ and $S(x)$ exhibit the minimal-area 
law (cf. below).
In Ref.~\cite{3}, the leading infra-red asymptotes of these Green functions have 
been obtained in terms of the lowest eigenvalues of respectively one- and two-body Hamiltonians with the linear potential.
Below, the full infra-red Green functions $G(x)$ and $S(x)$ will be found 
by means of a direct calculation of the corresponding path integrals with the minimal-area ans\"atze. 

\begin{figure}
\epsfig{file=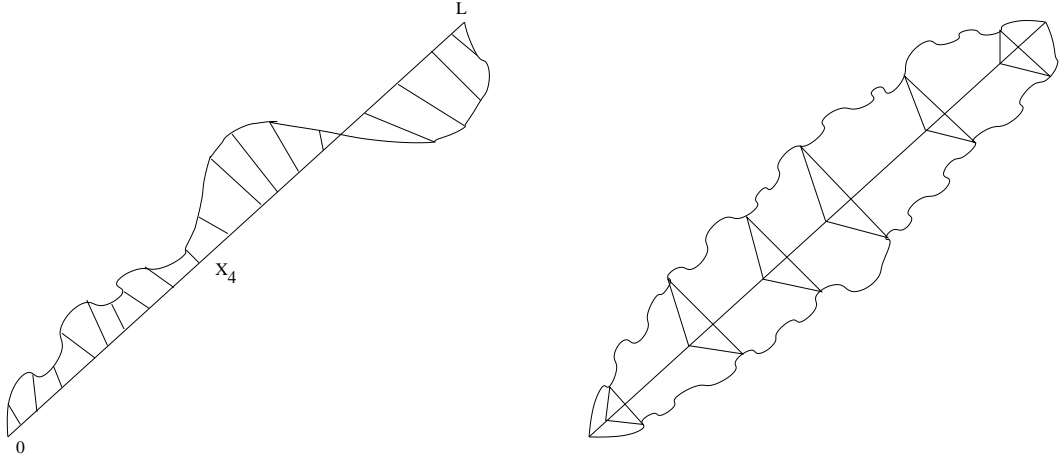, width=140mm}
\caption{\label{Fig12} A one-gluon gluelump (left) and a two-gluon gluelump (right). For illustration,
at various values of $x_4$, are also shown a straight-line adjoint string connecting the gluon to the static adjointly charged source (for the one-gluon gluelump) and three fundamental strings connecting two gluons and the static source (for the two-gluon gluelump).}
\end{figure}

We start with the Green function of a one-gluon gluelump,
\begin{equation}
\label{GG}
G(x)=\int_0^\infty ds\int_0^x{\cal D}z_\mu 
\exp\left(-\int_0^s\frac{\dot z_\mu^2}{4}d\lambda-
\sigma S_{\rm min}\right),
\end{equation}
where the minimal surface of area $S_{\rm min}$ is swept out by the adjoint string of tension $\sigma$, 
which connects the 
gluon to the adjointly charged source. While the effects of non-staticity of the source are considered in Appendix~A, here we assume the source to be static, i.e. evolving entirely along the 
$x_4$-axis. This means that only the $x_4$-coordinate of the point $x$ in Eq.~(\ref{GG}) is nonvanishing, i.e. 
$$x=({\bf 0},L).$$
In what follows, we continue with the notations $G(x)$ and $S(x)$, expressing 
these functions in terms of $L\equiv|x|$.

We calculate the path integral in Eq.~(\ref{GG}) by approximating $S_{\rm min}$ through 
the Cauchy--Schwarz inequality~\cite{4}:
\begin{equation}
\label{CS}
S_{\rm min}=\int_0^L d\tau|{\bf z}(\tau)|\le\sqrt{L\int_0^L d\tau{\bf z}^2}.
\end{equation}
It reduces the path integral to that of a harmonic oscillator with a variable frequency. Indeed, using the 
formula ${\rm e}^{-\sqrt{A}}=
\int_0^\infty\frac{d\lambda}{\sqrt{\pi\lambda}}\exp\left(-\lambda-\frac{A}{4\lambda}\right)$ 
with $A>0$, and 
changing the proper-time variable as $s\to\mu=\frac{L}{2s}$, we have 
\begin{equation}
\label{Gx}
G(x)\simeq\frac{L}{2}\int_0^\infty\frac{d\mu}{\mu^2}\int_0^L{\cal D}z_4\oint{\cal D}{\bf z}
\int_0^\infty\frac{d\lambda}{\sqrt{\pi\lambda}}\exp\left(-\lambda-\frac{\mu}{2}\int_0^Ld\tau\dot z_\mu^2-
\frac{\sigma^2L}{4\lambda}\int_0^Ld\tau{\bf z}^2\right),
\end{equation}
where $\oint{\cal D}{\bf z}$ denotes the integration over trajectories, 
which start and end up at ${\bf z}={\bf 0}$.
The kinetic term $\frac{\mu\dot z_\mu^2}{2}$ means, of course, that the auxiliary parameter $\mu$ can be 
viewed as an effective gluon mass. Furthermore, 
the approximate equality ``$\simeq$'' in Eq.~(\ref{Gx}) is understood in the sense of 
approaching the upper limit for $S_{\rm min}$ in Eq.~(\ref{CS}). We see then that the resulting  
path integral over $z_4(\tau)$ is that of a free particle,
$$
\int_0^L{\cal D}z_4\exp\left(-\frac{\mu}{2}\int_0^Ld\tau\dot z_4^2\right)=\sqrt{\frac{\mu}{2\pi L}}
{\rm e}^{-\mu L/2},$$
whereas the integral over ${\bf z}(\tau)$ is that of a harmonic oscillator. It reads 
\be
\label{78}
\oint{\cal D}{\bf z}
\exp\left(-\frac{\mu}{2}\int_0^L d\tau\dot {\bf z}^2-\frac{\sigma^2L}{4\lambda}\int_0^L d\tau{\bf z}^2\right)=
\left[\frac{\omega}{4\pi\sinh\left(\frac{\omega L}{2\mu}\right)}\right]^{3/2},
\ee
where $\omega\equiv\sigma\sqrt{2\mu L/\lambda}$ is the frequency of the oscillator.
Changing further the integration variable from 
$\lambda$ to $\xi\equiv\sigma L^{3/2}/\sqrt{2\mu\lambda}$, we can perform the $\mu$-integration:
$$
G(x)=\sigma\sqrt{\frac{L}{32\pi^5}}\int_0^\infty\frac{d\mu}{\sqrt{\mu}}
\int_0^\infty\frac{d\xi}{\sqrt{\xi}\sinh^{3/2}\xi}\exp\left(-\frac{\mu L}{2}-\frac{\sigma^2L^3}{2\mu\xi^2}\right)=$$
\be
\label{gg5}
=\frac{\sigma}{4\pi^2}\int_0^\infty\frac{d\xi}{\sqrt{\xi}\sinh^{3/2}\xi}{\rm e}^{-l/\xi},~~ 
{\rm where}~~ l\equiv\sigma L^2.
\ee
In the infra-red limit $l\gg 1$ of interest, this integral can be evaluated analytically as follows:
\be
\label{s9}
\int_0^1\frac{d\xi}{\xi^2}{\rm e}^{-l/\xi}+2^{3/2}\int_1^\infty\frac{d\xi}{\sqrt{\xi}}{\rm e}^{-\frac{3\xi}{2}-\frac{l}{\xi}}\simeq\frac{{\rm e}^{-l}}{l}+2^{3/2}\int_0^\infty\frac{d\xi}{\sqrt{\xi}}{\rm e}^{-\frac{3\xi}{2}-\frac{l}{\xi}}=
\frac{{\rm e}^{-l}}{l}+4\sqrt{\frac{\pi}{3}}{\rm e}^{-\sqrt{6\sigma}L}.
\ee
Here, the replacement of the 
lower limit of integration in the second integral by $0$ was legitimate, since the saddle-point value $\xi_{\rm s.p.}=\sqrt{2l/3}$, in the limit $l\gg 1$ at issue, is larger than $1$ (and therefore the contribution of the
integration region $0<\xi<1$ to the whole integral is exponentially small). Thus, the leading 
result in the large-$|x|$ limit reads
\begin{equation}
\label{B}
G(x)\simeq\frac{\sigma}{\sqrt{3\pi^3}}{\rm e}^{-\sqrt{6\sigma}|x|}.
\end{equation}
By means of Eq.~(\ref{D}), it yields the following function $D_1(x)$:
\begin{equation}
\label{d1}
D_1(x)=g^2C_{\rm f}\left(\frac{2\sigma}{\pi}\right)^{3/2}\frac{{\rm e}^{-\sqrt{6\sigma}|x|}}{|x|}.
\end{equation}
This expression can be compared with the one from Ref.~\cite{3}, 
\begin{equation}
\label{td1}
\tilde D_1(x)=g^2C_{\rm f}\frac{M_0\sigma}{2\pi}\cdot\frac{{\rm e}^{-M_0|x|}}{|x|},
\end{equation}
where the value $M_0\simeq 1.5{\,}{\rm GeV}$ was obtained from the Schr\"odinger 
equation with the linear potential. 
To this end, we evaluate the adjoint string tension via the so-called Casimir-scaling 
hypothesis~\cite{CasHyp}.
This hypothesis, supported both by lattice simulations~\cite{CasLat} and analytic studies~\cite{rev2,Cas},
suggests proportionality of the string tension, in a given representation of SU($N_c$), to the 
quadratic Casimir operator of that representation. For the adjoint representation of SU(3), it yields 
$\sigma=\frac94\sigma_{\rm f}$, where the value of the string tension in the fundamental representation is 
$\sigma_{\rm f}=(440{\,}{\rm MeV})^2$.  
Accordingly, the obtained 
\be
\label{m1}
{\rm mass}~ {\rm of}~ {\rm the}~ 1g~ {\rm  gluelump}=
\sqrt{6\sigma}\simeq1.6{\,}{\rm GeV},
\ee 
turns out to be 
very close to the quoted value of $M_0$.
Note that, in principle, a gluon is confined up to arbitrarily large distances only at $N_c\gg 1$, whereas
at $N_c\sim 1$ it can be screened by other gluons. In the large-$N_c$ limit,
$\sigma\to 2\sigma_{\rm f}$, and the exponents in Eqs.~(\ref{d1}) and (\ref{td1}) coincide numerically extremely well, since in that case $\sqrt{6\sigma}\simeq1.5{\,}{\rm GeV}$.

We proceed now to the Green function of the two-gluon gluelump:
$$S(x)=\frac{L^2}{4}\int_0^\infty \frac{d\mu}{\mu^2}\int_0^\infty \frac{d\bar\mu}{\bar\mu^2}
\int_0^x{\cal D}z_\mu\int_0^x{\cal D}\bar z_\mu
\exp\Biggl\{-\frac{\mu}{2}\int_0^L d\tau\dot z_\mu^2-\frac{\bar\mu}{2}\int_0^L d\tau\dot{\bar z}_\mu^2-$$
$$-\sigma_{\rm f}\int_0^L d\tau\left(|{\bf z}|+|\bar{\bf z}|+|{\bf z}-\bar{\bf z}|\right)\Biggr\}.$$
Trying out the Cauchy--Schwarz inequality, Eq.~(\ref{CS}), for each of the three distances separately would involve integrations over three auxiliary parameters. In order to reduce this number to just one, it is more useful to apply the 
Cauchy--Schwarz inequality in the form
$$\frac1n\sum\limits_{i=1}^{n}A_i\le\sqrt{\frac1n\sum\limits_{i=1}^{n}A_i^2},~~~ {\rm where}~~ A_i>0.$$
At $n=3$, it yields
$$|{\bf z}|+|\bar{\bf z}|+|{\bf z}-\bar{\bf z}|\le \sqrt{3}\cdot\sqrt{{\bf z}^2+\bar{\bf z}^2+({\bf z}-\bar{\bf z})^2}.$$
Now, so long as the common square root for the three distances is assembled, we can again apply the Cauchy--Schwarz inequality in the form of Eq.~(\ref{CS}) with only one auxiliary integration:
$$S(x)\simeq\frac{L^2}{4}\int_0^\infty \frac{d\mu}{\mu^2}\int_0^\infty \frac{d\bar\mu}{\bar\mu^2}
\int_0^x{\cal D}z_\mu\int_0^x{\cal D}\bar z_\mu\int_0^\infty\frac{d\lambda}{\sqrt{\pi\lambda}}
\exp\Biggl\{-\lambda-\frac{\mu}{2}\int_0^L d\tau\dot z_\mu^2-\frac{\bar\mu}{2}\int_0^L d\tau\dot{\bar z}_\mu^2-$$
\begin{equation}
\label{S}
-\frac{3\sigma_{\rm f}^2L}{4\lambda}\int_0^L d\tau\left[{\bf z}^2+\bar{\bf z}^2+({\bf z}-\bar{\bf z})^2\right]\Biggr\}.
\end{equation}
Similar to the one-gluon gluelump, integrations over $z_4(\tau)$ and $\bar z_4(\tau)$ in this formula yield
$$\int_0^L{\cal D}z_4\int_0^L{\cal D}\bar z_4\exp\left[-\frac12\int_0^L d\tau(\mu\dot z_4^2+\bar\mu\dot{\bar z}_4^2)\right]=
\frac{\sqrt{\mu\bar\mu}}{2\pi L}\exp\left[-\frac{(\mu+\bar\mu)L}{2}\right].$$

\begin{figure}
\epsfig{file=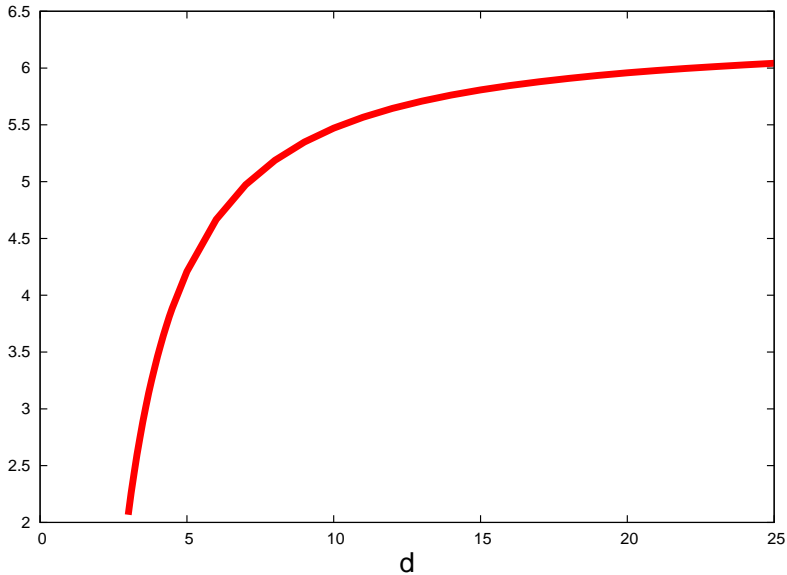, width=110mm}
\caption{\label{xd}Plotted is $-\frac{\ln(S(x)/\sigma_{\rm f}^2)}{d}$ in the range $d\in[3,25]$.}
\end{figure}

We observe now that, if the terms ${\bf z}^2$ and $\bar{\bf z}^2$ in Eq.~(\ref{S}) were absent, the path integral would be that of two (mutually interacting but otherwise free) particles, which was calculated in Ref.~\cite{4}.
Here, however, we have to deal with two (also mutually interacting) harmonic oscillators. Nevertheless, 
such a path integral $\oint{\cal D}{\bf z}\oint{\cal D}\bar{\bf z}$ can still be calculated. This 
turns out to be possible upon the diagonalization of the action by virtue of the known fact that two positively
definite quadratic forms (that are, the kinetic and the potential energies) can be diagonalized simultaneously. 
Passing to the new integration trajectories ${\bf u}(\tau)$ and ${\bf v}(\tau)$ according to the formulae
\begin{equation}
\label{ch}
{\bf z}={\bf u}+\alpha{\bf v},~~ \bar{\bf z}=\beta{\bf u}+{\bf v},
\end{equation}
we obtain the diagonalization conditions
$$\mu\alpha+\bar\mu\beta=0,~~ \alpha+\beta=(\alpha-1)(\beta-1).$$
The solution to these equations is straightforward:
\begin{equation}
\label{ab}
\beta(\mu,\bar\mu)=-\frac{\mu}{\bar\mu}\cdot\alpha(\mu,\bar\mu),~~~~ {\rm where}~~  \alpha(\mu,\bar\mu)=1-\frac{\bar\mu}{\mu}+\sqrt{\frac{\bar\mu}{\mu}+\left(1-\frac{\bar\mu}{\mu}
\right)^2},
\end{equation}
where we have chosen for concreteness the ``$+$'' sign in front of the last square root.
Then the kinetic- and the potential-energy terms read
$$\mu\dot{\bf z}^2+\bar\mu\dot{\bar{\bf z}}^2=
(\mu+\bar\mu\beta^2)\dot{\bf u}^2+(\mu\alpha^2+\bar\mu)\dot{\bf v}^2,$$
$${\bf z}^2+\bar{\bf z}^2+({\bf z}-\bar{\bf z})^2=2\left[(\beta^2-\beta+1){\bf u}^2+(\alpha^2-\alpha+1){\bf v}^2\right].$$
The Jacobian stemming from the change of integration trajectories in Eq.~(\ref{ch}) is, of course, also $\mu$- and $\bar\mu$-dependent, namely
$${\cal D}{\bf z}{\,}{\cal D}\bar{\bf z}=|1-\alpha\beta|{\,}{\cal D}{\bf u}{\,}{\cal D}{\bf v}=
\left(1+\alpha^2\frac{\mu}{\bar\mu}\right){\cal D}{\bf u}{\,}{\cal D}{\bf v}.$$
Thus, we have reduced the path integral $\oint{\cal D}{\bf z}\oint{\cal D}\bar{\bf z}$ to the product of two well-known path integrals for non-interacting harmonic oscillators. 
Introducing finally the dimensionless variables
$$\nu\equiv\frac{\mu}{\sqrt{\sigma_{\rm f}}},~~~ \bar\nu=\frac{\bar\mu}{\sqrt{\sigma_{\rm f}}},~~~ 
d\equiv\sqrt{\sigma_{\rm f}}L,$$
we can write down the result in the following form:
$$S(x)=\frac{3^{3/2}\sigma_{\rm f}^2}{64\pi^{9/2}}\cdot d^{5/2}\int_0^\infty\frac{d\nu}{\nu^{3/2}}\int_0^\infty\frac{d\bar\nu}{\bar\nu^{3/2}}
\left(1+\alpha^2\frac{\nu}{\bar\nu}\right)\int_0^\infty\frac{d\lambda}{\lambda^2}
\exp\left[-\lambda-\frac{(\nu+\bar\nu)d}{2}\right]\times$$
$$\times\left[(\beta^2-\beta+1)(\alpha^2-\alpha+1)(\nu+\bar\nu\beta^2)(\bar\nu+\nu\alpha^2)\right]^{3/4}\times$$
$$\times\sinh^{-3/2}\left( d^{3/2}\sqrt{\frac{3}{\lambda}\cdot\frac{\beta^2-\beta+1}{\nu+\bar\nu\beta^2}}\right)\cdot\sinh^{-3/2}\left(d^{3/2} \sqrt{\frac{3}{\lambda}\cdot\frac{\alpha^2-\alpha+1}{\bar\nu+\nu\alpha^2}}\right),$$
where $\alpha\equiv\alpha(\nu,\bar\nu)$ and $\beta\equiv\beta(\nu,\bar\nu)$ are given by Eq.~(\ref{ab}).
The remaining ordinary integrations in this formula have been performed numerically. 
In Fig.~\ref{xd}, we plot the so-calculated $-\frac{\ln(S(x)/\sigma_{\rm f}^2)}{d}$ as a 
function of $d$ in the range $d\in[3,25]$, and observe an asymptotic approach of this quantity to 6.0
at large $d$. For $\sigma_{\rm f}=(440{\,}{\rm MeV})^2$, the corresponding 
\be
\label{m2}
{\rm mass}~ {\rm of}~ {\rm the}~ 2g~ {\rm  gluelump}=
6\sqrt{\sigma_{\rm f}}\simeq2.6{\,}{\rm GeV},
\ee
agrees remarkably well with its value of 2.56~GeV
found in Ref.~\cite{3} within the Hamiltonian approach.

\section{Concluding remarks}
In this paper, we have analytically confirmed the statement that
nonperturbative-nonconfining and confining self-interactions of stochastic background Yang--Mills fields
can have different correlation lengths. Namely, we have obtained these lengths 
from the Green functions of one- and 
two-gluon gluelumps, by means 
of an analytic treatment of the quantum-mechanical path integrals for one and two gluons (inter)connected
by strings with the static adjoint source. The resulting inverse correlation 
lengths are given by Eqs.~(\ref{m1}) and (\ref{m2}). They turn out to be in a remarkably good agreement
with those found in Ref.~\cite{3} by means of a different, Hamiltonian, approach. We have also demonstrated 
that, in accordance with the lattice results~\cite{dm1}, the two-point correlation function of nonperturbative-nonconfining background fields decreases with the deformation of the path interconnecting these 
two points.
On the technical side, for the calculation of the Green function of a two-gluon gluelump, a novel method has been developed of an analytic
evaluation of the path integral with the common minimal surface formed by two gluons together with the 
adjointly charged source.

In the forthcoming studies, we plan to incorporate dynamical quarks in this theoretical framework.
Their appearance can lead to the breaking of not only fundamental strings, but also of the adjoint ones --- through the 
creation of $\bar q$-$g$-$q$ hybrids. An effective reduction, due to the string-breaking, of string 
world sheets in the spatial directions can lead to their enhancement in the temporal direction, 
that is, to the increase of the vacuum correlation lengths. In this way, vacuum correlation lengths in full
QCD can exceed those obtained here in pure Yang--Mills theory by a factor of 3, as 
suggested by the lattice simulations~\cite{lat1} and the 
phenomenology of hadronic collisions~\cite{b}.

\begin{acknowledgments}
The author is grateful to J.E.F.T.~Ribeiro for the very useful discussions. He has also benefitted from numerous discussions with N.~Brambilla, A.~Di~Giacomo, H.G.~Dosch, E.~Meggiolaro, M.G.~Schmidt, Yu.A.~Simonov, and A.~Vairo. 
This work was supported by the Portuguese Foundation for Science and Technology
(FCT, program Ci\^encia-2008) and by 
the Center for Physics of Fundamental Interactions (CFIF) at Instituto Superior
T\'ecnico (IST), Lisbon. 
\end{acknowledgments}

\appendix

\section{Path-dependence of the function $D_1(x)$}
Lattice simulations~\cite{dm1} indicate that, if the path interconnecting the points $x$ and $x'$ in Eq.~(\ref{09}) deviates from the straight-line one, the correlation function decreases.
In this Appendix, we test this indication analytically, at an 
example of the function $D_1(x)$, by evaluating it for paths of various shapes. The approach we use for this 
study is the same as the one used in Section~II. 

We parametrize the path interconnecting $x$ and $x'$ by some
vector-function ${\bf w}(\tau)$, whose concrete form will be specified below. Accordingly, the minimal-area
ansatz~(\ref{CS}) gets modified, and takes the form
$$S_{\rm min}=\int_0^L d\tau|{\bf z}(\tau)-{\bf w}(\tau)|\le\sqrt{L\int_0^L d\tau({\bf z}-{\bf w})^2}.$$ 
The emerging path integral $\oint{\cal D}{\bf z}$ turns out to be similar to that of Eq.~(\ref{78}), and reads (cf. Ref.~\cite{fh})
$$\oint{\cal D}{\bf z}
\exp\left(-\frac{\mu}{2}\int_0^L d\tau\dot {\bf z}^2-\frac{\sigma^2L}{4\lambda}\int_0^L d\tau{\bf z}^2
+\frac{\sigma^2L}{2\lambda}\int_0^L d\tau{\bf z}{\bf w}\right)=
\left[\frac{\omega}{4\pi\sinh\left(\frac{\omega L}{2\mu}\right)}\right]^{3/2}\times$$
$$\times\exp\left\{\frac{\omega^3 L^2}{8\mu^2\sinh\left(\frac{\omega L}{2\mu}\right)}
\int_0^1 du\int_0^u dv\sinh\left[\frac{\omega L}{2\mu}(1-u)\right]\sinh\left(\frac{\omega L}{2\mu}v\right)
{\bf w}(Lu){\bf w}(Lv)\right\}.$$
The corresponding path-dependent Green function $G_{\bf w}(x)$ generalizes Green function $G(x)$, and can be written as [cf. Eq.~(\ref{gg5})]
\be
\label{a1}
G_{\bf w}(x)=\sigma\sqrt{\frac{L}{32\pi^5}}\int_0^\infty\frac{d\mu}{\sqrt{\mu}}
\int_0^\infty\frac{d\xi}{\sqrt{\xi}\sinh^{3/2}\xi}\exp\left\{-
\frac{\mu L}{2}[1+f(\xi)]-\frac{\sigma^2L^3}{2\mu\xi^2}\right\}.
\ee
The function 
\be
\label{a2}
f(\xi)\equiv\frac{\xi^2}{L^2}\left\{\int_0^1 du{\bf w}^2(Lu)-\frac{2\xi}{\sinh\xi}
\int_0^1 du\int_0^u dv\sinh[\xi(1-u)]\sinh(\xi v)
{\bf w}(Lu){\bf w}(Lv)\right\},
\ee
introduced here, vanishes in the limiting case ${\bf w}={\bf 0}$.
Performing again the $\mu$-integration exactly, and using Eq.~(\ref{D}) with $G(x)$ replaced by 
$G_{\bf w}(x)$, we obtain the path-dependent correlation function
\be
\label{a5}
D_1^{({\bf w})}(x)=\sigma^2{\,}\frac{g^2C_{\rm f}}{\pi^2}\int_0^\infty\frac{d\xi}{(\xi\sinh\xi)^{3/2}}
{\rm e}^{-\frac{l}{\xi}\sqrt{1+f(\xi)}},
\ee
where $l$ is defined in Eq.~(\ref{gg5}).
To calculate this integral, we parametrize the path ${\bf w}(\tau)$ in the form which provides 
smooth approximations to step-like paths used in Ref.~\cite{dm1}. Namely, we consider two types of smooth paths: 
$${\bf w}^{(1)}(\tau)=\frac{L}{2}\left(\sin\frac{\pi\tau}{L},0,0\right),~~~ 
{\bf w}^{(2)}(\tau)=\frac{L}{2}\left(\sin\frac{2\pi\tau}{L},0,0\right).$$
Each of these paths lies in the (1,4)-plane, deviating from the $x_4$-axis to the maximum distance $L/2$, which is still compatible with $L$. In general, for larger deviations, one can expect cancellations of contributions 
stemming from the mutually backtracking pieces of the path. 

\begin{figure}
\psfrag{x}{\Large{$\xi$}}
\psfrag{y}{\Large{$f(\xi)$}}
\epsfig{file=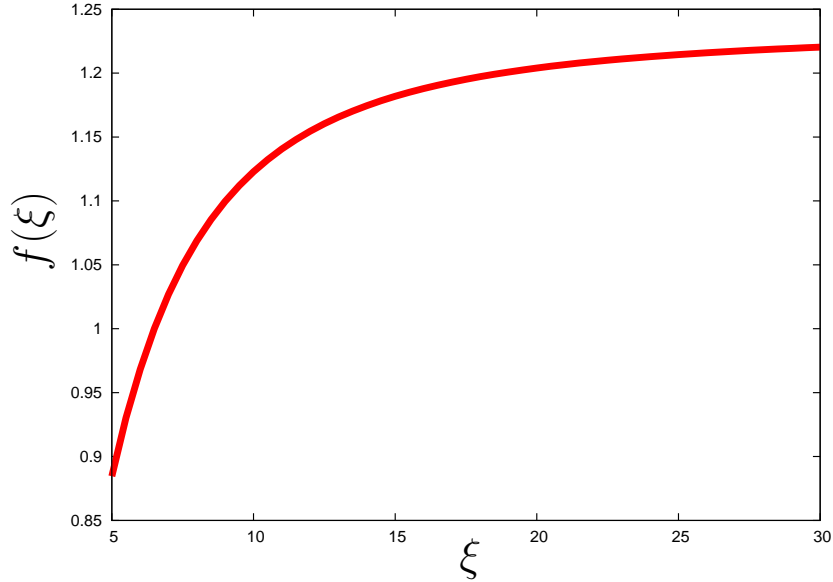, width=110mm}
\caption{\label{WW}Function~(\ref{a10}) for $\xi\gg 1$.}
\end{figure}

Consider first 
the path ${\bf w}^{(1)}(\tau)$. The corresponding function~(\ref{a2}) reads
$$
f(\xi)=\frac{\xi^2}{4}\left\{\frac12-\frac{2\xi}{\sinh\xi}
\int_0^1 du\int_0^u dv\sinh[\xi(1-u)]\sinh(\xi v)\sin(\pi u)\sin(\pi v)\right\}.
$$
The $\xi$-integration in Eq.~(\ref{a5}) can be performed in the same way as in Eq.~(\ref{s9}), by splitting the integration region into the intervals $[0,1]$ and $(1,\infty)$. In the interval $[0,1]$, 
$f(\xi)=\frac{\xi^2}{8}+{\cal O}(\xi^4)$, and it can be disregarded altogether compared to 1. The 
contribution stemming from this interval again appears exponentially suppressed compared to the leading  
contribution stemming from the interval $(1,\infty)$. For $\xi\in(1,\infty)$, the exponential part of the 
$\xi$-dependence of the integrand, which determines the position of the saddle-point, reads [cf. Eq.~(\ref{s9})] 
${\rm e}^{-\frac{3\xi}{2}-\frac{l}{\xi}\sqrt{1+f(\xi)}}$. To figure out the extent to which the function 
$f(\xi)$ affects the saddle point, one needs to consider the corresponding saddle-point 
equation, 
\be
\label{a7}
\frac32-\frac{l}{\xi^2}\sqrt{1+f(\xi)}+\frac{l}{2\xi}{\,}\frac{f'(\xi)}{\sqrt{1+f(\xi)}}=0,
\ee
where 
\be
\label{a10}
f(\xi)\simeq\frac{\xi^2}{4}\left[\frac12-4\xi{\rm e}^{-\xi}
\int_0^1 du\int_0^u dv\sinh[\xi(1-u)]\sinh(\xi v)\sin(\pi u)\sin(\pi v)\right]~~ {\rm at}~~ \xi>1.
\ee
One can then prove numerically that the absolute value of the second term 
on the left-hand side of Eq.~(\ref{a7}) exceeds the third term by at least one order of magnitude.
Therefore, the third term can be disregarded compared to the second one, for the reason of 
smallness of $f'(\xi)$. Accordingly, approximating $f(\xi)$ by some constant $f$, we obtain 
the saddle-point value $\xi_{\rm s.p.}=\sqrt{2l\sqrt{1+f}/3}$, which is much larger than 1 in the limit $\sqrt{l}\gg 1$ of interest. And indeed, for $\xi\gg 1$, one can see a very weak variation of the function $f(\xi)$, which is illustrated by Fig.~\ref{WW}, where this function is plotted in the range $\xi\in[5,30]$.
We therefore approximate $f(\xi)$ by its limiting value at $\xi\gg 1$, i.e. set $f\simeq1.23$.

Then, given that $\xi_{\rm s.p.}$ lies deeply inside the region $(1,\infty)$, we can replace the 
lower limit of integration by 0 [cf. Eq.~(\ref{s9})], and obtain
\be
\label{kl}
D_1^{({\bf w})}(x)\simeq\sigma^2{\,}\frac{g^2C_{\rm f}}{\pi^2}\cdot 2^{3/2}\int_0^\infty
\frac{d\xi}{\xi^{3/2}}{\rm e}^{-\frac{3\xi}{2}-\frac{l}{\xi}\sqrt{1+f}}=
g^2C_{\rm f}\left(\frac{2\sigma}{\pi}\right)^{3/2}
\frac{{\rm e}^{-\sqrt{6\sigma}\cdot\sqrt[4]{1+f}\cdot|x|}}{\sqrt[4]{1+f}\cdot|x|}.
\ee
Comparing this expression with Eq.~(\ref{d1}), we observe a decrease of the amplitude of the correlation 
function by a factor of $\sqrt[4]{1+f}\simeq1.22$, in a qualitative agreement with indications 
of Ref.~\cite{dm1} on the dominance of the straight-line path. Furthermore, we obtain also
an increase of the mass of the one-gluon gluelump, 
Eq.~(\ref{m1}), by the same factor of $1.22$. Such an increase yields an even stronger suppression of  
contributions produced by curved paths.

In general, the present approach yields, for a given path, a decrease of the vacuum correlation length and 
the amplitude of the correlation function 
by the same amount as compared to these quantities for the straight-line path. With the 
deformation of the path, the suppression factor increases. This statement can be illustrated by considering 
the path 
${\bf w}^{(2)}(\tau)$, which is deformed stronger than ${\bf w}^{(1)}(\tau)$. The corresponding function
$$f(\xi)\simeq\frac{\xi^2}{4}\left[\frac12-4\xi{\rm e}^{-\xi}
\int_0^1 du\int_0^u dv\sinh[\xi(1-u)]\sinh(\xi v)\sin(2\pi u)\sin(2\pi v)\right]~~ {\rm at}~~ \xi>1$$
also exhibits a rapid levelling-off, similar to its counterpart given by Eq.~(\ref{a10}) [cf. Fig.~\ref{WW}]. The difference is that the limiting value of the function $f(\xi)$ at $\xi\gg 1$ appears in this case larger, namely 
$f\simeq4.86$, instead of $1.23$. Accordingly, the suppression factor in Eq.~(\ref{kl}) becomes  $\sqrt[4]{1+f}\simeq1.56$,
instead of $1.22$. Such an increase of the factor $\sqrt[4]{1+f}$ is quite fast, in spite of its slow, 
fourth-root, functional dependence. Thus, our analysis suggests a substantial suppression of contributions 
to the two-point correlation function, which stem from strongly deformed paths.

\end{document}